\documentclass[aps,preprint]{revtex4}%
\usepackage{amsfonts}
\usepackage{amsmath}
\usepackage{amssymb}
\usepackage{graphicx}%
\begin{document}
\title[ ]{Classical and Quantum Interpretations Regarding Thermal Behavior in a
Coordinate Frame Accelerating Through Zero-Point Radiation}
\author{Timothy H. Boyer}
\affiliation{Department of Physics, City College of the City University of New York, New
York, New York 10031}
\keywords{}
\pacs{}

\begin{abstract}
A relativistic classical field theory with zero-point radiation involves a
vacuum corresponding to a scale-invariant spectrum of random classical
radiation in spacetime with the overall constant chosen to give an energy
$(1/2)\hbar\omega$ per normal mode in inertial frames. \ Classical field
theory with classical zero-point radiation gives the same field correlation
functions as quantum field theory for the symmetrized products of the
corresponding free massless fields in inertial frames; however, the
interpretations in classical and quantum theories are quite different.
\ Quantum field theory has photons in thermal radiation but not in the vacuum
state; classical theory has radiation in both situations. \ The contrast in
interpretations is most striking for the Rindler coordinate frame accelerating
through zero-point radiation; classical theory continues tensor behavior over
to the Rindler frame, whereas quantum theory introduces a new Rindler vacuum
state. \ The classical interpretation of thermal behavior rests on two
fundamental principles. i) A scale-invariant distribution of random radiation
cannot correspond to thermal radiation at non-zero temperature. ii) A
scale-invariant distribution of random radiation can acquire a correlation
time which reflects the parameters of a spacetime trajectory through the
scale-invariant radiation. \ Based on these principles, classical theory finds
no basis for an accelerating observer to reinterpret zero-point radiation in
terms of thermal radiation. \ In contrast, quantum field theory claims that an
observer uniformly accelerated through zero-point flucturations of the
Minkowski vacuum encounters a thermal bath at the temperature $T=\hbar a/(2\pi
ck_{B}).$

\end{abstract}
\maketitle

\section{Introduction}

In the 1970's in connection with Hawking radiation from a black hole,
Davies,\cite{Davies} Unruh,\cite{Unruh} and Fulling\cite{Fulling} suggested
the "thermal effects of acceleration." \ Thus it was\ noted that the two-field
vacuum correlation function in time for a scalar field in a Rindler coordinate
frame accelerating through the Minkowski vacuum involved the same Planck
distribution as is found for thermal radiation in an inertial frame. \ Indeed,
the quantum aspects of radiation viewed in accelerating frames have been
developed extensively under the heading of "the Unruh effect," and a recent
review article by Crispino, Higuchi, and Matsas\cite{Crispino} lists hundreds
of references on the subject. Their review states,\cite{Crispino2} "... the
Unruh effect expresses the fact that uniformly accelerated observers in
Minkowski space-time ... associate a thermal bath of Rindler particles ... to
the no-particle state of inertial observers ..." \ The appearance of thermal
behavior from basic aspects of quantum field theory has intrigued many
physicists, and Sciama\cite{Sciama} has proposed that we may be poised for a
new synthesis of some fundamental aspects of physics. \ In the present
article, we wish to sharpen our understanding of the "thermal effects of
acceleration" by highlighting the contrasting interpretations provided by
classical and quantum field theories.

Since quantum theory developed out of classical theory, we expect strong
connections between classical and quantum theories. Indeed for free fields and
linear oscillator systems in inertial frames, there is a general connection
between quantum theory and classical electrodynamics with classical
electromagnetic zero-point radiation (stochastic electrodynamics). \ In 1975
it was shown\cite{1975agree} that for free fields and linear systems in
inertial frames, the classical theory with zero-point radiation gives average
values which are in exact agreement with the expectation values of the
symmetrized operator products for the corresponding quantum systems, both in
the vacuum and also in thermal equilibrium at non-zero temperature.\ \ Because
of this agreement, certain aspects of physics, such as the fluctuation aspects
of thermal radiation,\cite{1969} can be understood alternatively in terms of
quanta or in terms of fluctuations of classical radiation including classical
zero-point radiation. The physical interpretations given for the results are,
however, strikingly different. The classical theory regards zero-point
radiation and thermal radiation as alike in character, with finite temperature
involving a finite density of classical radiation above the classical
zero-point radiation. \ In contrast, the quantum theory regards the vacuum
state as involving fluctuations (including correlations in these fluctuations)
but no energy quanta, while thermal radiation involves a characteristic
distribution of radiation quanta.

In the early 1980's, the close connection between classical and quantum
theories for linear systems was applied to show that classical field theories
with classical zero-point radiation showed some of the same "thermal effects
of acceleration" as were found in quantum field theory.\cite{thermBC}
\ Although the quantum analysis of the Unruh effect has flourished in recent
years, the classical perspective has languished.\ Nevertheless, it seems wise
for physicists to be aware of the areas of agreement and disagreement between
the classical and quantum interpretations. In the present article we wish to
contrast the classical and quantum perspectives regarding the "thermal effects
of acceleration" through the vacuum. \ Our comparison will involve only
\textit{massless} free \textit{scalar} fields, leaving the electromagnetic
case and linear oscillator systems for future work. \ The comparison shows
contradictory physical interpretations between classical and quantum theories.
\ Whereas classical physics finds only zero-point radiation on acceleration
through the vacuum, the quantum literature claims that acceleration through
the vacuum provides a "thermal bath."

The outline of the presentation is as follows. \ In Section II we give a
cursory summary of classical electrodynamics with classical electromagnetic
zero-point radiation. \ Then we turn to the massless scalar field and discuss
the idea of zero-point radiation as the scale-invariant spectrum of random
classical radiation in a general spacetime. \ Next it is pointed out that
thermal radiation involves a finite density of radiation above the vacuum
state. \ The finite thermal density must be associated with radiation
correlation lengths and correlation times. \ In Section III, we review the
general connection between classical and quantum free fields in Minkowski
spacetime which was noted in 1975. We obtain the correlation functions for the
fields, both for zero-point radiation and for thermal radiation where the
quantum analysis finds the presence of thermal photons. \ In Section IV, we
introduce the Rindler coordinate frame. We insert Rindler coordinates into the
two-point field correlation function found for the Minkowski vacuum. \ At a
single spatial coordinate point but at different times, the correlation
function corresponds to the Planck spectrum; at a single time but at different
spatial points, the correlation function corresponds to zero-point radiation.
\ Since there is no correlation length for the classical radiation at a single
time, we conclude in classical physics that there is no thermal radiation
present. \ Indeed, the correlation time is related to the parameters of the
coordinate trajectory through spacetime and not to a thermodynamic ensemble.
\ On the other hand, quantum field theory uses the canonical ensemble as its
criterion for thermal behavior and declares that the accelerating coordinate
system experiences a thermal bath. \ Since the classical situation involves no
spatial correlation length, no energy above the zero-point radiation, and no
characteristic "sloshing" for the zero-point radiation in an accelerating box,
classical physics does not find a "thermal bath" in the Rindler frame.
\ Finally in Section V, we give a closing summary.

\section{Classical Field Theory with Zero--Point Radiation in a General
Spacetime}

\subsection{Summary of Classical Electron Theory with Classical
Electromagnetic Zero-Point Radiation (Stochastic Electrodynamics)}

Classical electron theory involves the interactions of classical point charges
with electromagnetic fields. \ The theory requires a choice of homogeneous
boundary condition on Maxwell's equations for the electromagnetic fields.
\ The traditional classical electron theory of H. A. Lorentz chooses this
homogeneous boundary condition to correspond to zero; all radiation arises at
a finite time from the acceleration of charged particles. \ This traditional
theory provides classical descriptions of a number of microscopic phenomena,
such as optical dispersion, Faraday rotation, and the normal Zeeman
effect.\cite{Lorentz} \ \ However, a far better choice of boundary condition
assumes that the homogeneous boundary condition on Maxwell's equations
corresponds to random classical electromagnetic radiation with a
Lorentz-invariant spectrum, classical electromagnetic zero-point
radiation.\cite{reviews} \ This classical electron theory with classical
electromagnetic zero-point radiation is often termed "stochastic
electrodynamics." \ At present, stochastic electrodynamics provides the best
\textit{classical} description of microscopic physical phenomena. The
inclusion of classical electromagnetic zero-point radiation extends the
descriptive power of classical electron theory to Casimir forces, van der
Waals forces, diamagnetism, specific heats of solids, blackbody
radiation,\cite{reviews} and the ground state of hydrogen,\cite{ColeH} all of
which can be described in terms of linear systems or Coulomb potentials. The
one unknown scale factor for the zero-point radiation is chosen to give
numerical agreement with the experimentally observed Casimir forces, and the
numerical value is immediately recognized as corresponding to Planck's
constant $\hbar$. Although the limits of applicability of the classical theory
are still not known, the theory disagrees with quantum theory for non-Coulomb,
nonrelativistic nonlinear potentials.\cite{reviews}

\subsection{Classical Zero-Point Radiation in a General Spacetime}

Although classical electromagnetic zero-point radiation was originally derived
based upon Lorentz-invariance in Minkowski spacetime,\cite{reviews} zero-point
radiation can also be characterized as the $\sigma_{ltU^{-1}}$-scale invariant
spectrum of homogeneous, isotropic, random classical radiation.\cite{scaling}
\ Thus if the standards for measurement of length, time, and energy are
changed simultaneously $l\rightarrow l^{\prime}=\sigma l,$ $t\rightarrow
t^{\prime}=\sigma t,$ and $U\rightarrow U^{\prime}=U/\sigma,$ then the
spectrum of zero-point radiation is unchanged, as are the values of the speed
of light in vacuum $c$, the charge of the electron $e,$ and the value of
Planck's constant $\hbar.$ \ Indeed invariance under this scale transformation
uniquely determines the spectrum of classical zero-point radiation up to one
over-all multiplicative constant. \ This scale invariance is a natural
assumption so as to avoid introducing an intrinsic length into the
field-theory vacuum. \ It is also a natural invariance which is expected to
hold in a general spacetime.

In order to simplify the analysis presented here, we will turn from the
electromagnetic field theory over to the theory of a massless \textit{scalar}
field $\phi.$ \ Within Minkowski spacetime, the field satisfies the scalar
wave equation%
\begin{equation}
\nabla^{2}\phi-\frac{1}{c^{2}}\frac{\partial^{2}\phi}{\partial t^{2}}=0
\end{equation}
and has an energy in the field given by
\begin{equation}
U=\int d^{3}x\frac{1}{8\pi}\left[  \frac{1}{c^{2}}\left(  \frac{\partial\phi
}{\partial t}\right)  ^{2}+\left(  \nabla\phi\right)  ^{2}\right]
\end{equation}
For the vacuum, we expect that the two-point field correlation function should
involve the scale of zero-point radiation $\hbar$ and the speed of light in
vacuum $c.$ \ Dimensional analysis based on the energy equation (2) for
Minkowski spacetime indicates that the two-point field correlation for the
free vacuum field $\phi_{0}$\ at spacetime points $P$ and $Q$ in a general
spacetime must take the form%
\begin{equation}
\left\langle \phi_{0}(P)\phi_{0}(Q)\right\rangle =const\times\hbar
c/(length)^{2}%
\end{equation}
In order to maintain the covariance of the expression, the length must involve
the distance along a geodesic curve between the spacetime points $P$ and $Q$.
\ If we normalize to the expression which in Minkowski spacetime corresponds
to an energy $(1/2)\hbar\omega$ per normal mode,\cite{conformal} we have
finally the correlation function corresponding to classical zero-point
radiation in a general spacetime%
\begin{equation}
\left\langle \phi_{0}(P)\phi_{0}(Q)\right\rangle =\frac{-\hbar c}%
{\pi(\text{distance along a geodesic curve from }P\text{ to }Q)^{2}}%
\end{equation}
where the overall sign is chosen to correspond to the metric signature
$(+,-,-,-)$. \ We see that this classical zero-point radiation vacuum
situation is homogeneous, isotropic, and $\sigma_{ltU^{-1}}$-scale invariant
in a general spacetime. \ There is no correlation length or correlation time
associated with the zero-point radiation and the total energy density of
zero-point radiation is divergent.

\subsection{Fundamental Aspects of Thermal Radiation}

Thermal radiation involves a finite spatial energy density $u(T)$ above the
vacuum situation. \ The presence of a finite density of radiation means that
there must be a preferred coordinate frame; the coordinate-independent form
found for the vacuum situation in Eq. (4) is no longer possible. \ Indeed, it
is a familiar idea that thermal radiation equilibrium at a finite non-zero
temperature requires a confining box which determines a preferred coordinate
frame for the thermal radiation. \ The time evolution of the random radiation
is given by the normal-mode behavior in time. \ In the coordinate frame at
rest with respect to the box, \ the thermal correlation function will involve
a finite correlation length $\lambda_{T}$ and a finite correlation time
$\lambda_{T}/c$ associated with the finite density $u_{T}$ of thermal
radiation. The correlation length $\lambda_{T}$ will be associated with the
wavelength of the waves where the thermal energy per normal mode is comparable
to the zero-point energy per normal mode. \ Since the spectrum involves only a
finite total thermal energy $\mathcal{U}_{T}$ in a finite box, the radiation
energy per normal mode $U_{T}(\omega,T)$ must decrease at high frequencies
(short wavelengths). Thus thermal radiation at finite non-zero temperature
must involve radiation modes which are distinguished based upon the connection
between energy and frequency or wavelength. \ The correlation length
$\lambda_{T}$ \ or correlation time $t_{T}=\lambda_{T}/c=2\pi/\omega_{T}$
(corresponding to the transition mode between thermal energy and zero-point
energy) is exactly the parameter which appears in the Wien displacement
theorem $T\lambda_{T}=const$ for thermal radiation. The contrast between the
vacuum situation and the thermal situation for finite temperature $T>0$ is
thus quite clear. The zero-point radiation of the vacuum given in Eq. (4) has
no correlation length or correlation time associated with radiation, just as
there is no finite energy density in the vacuum. On the other hand, thermal
radiation indeed has a finite correlation length $\lambda_{T}$ and correlation
time $t_{T}$ which is associated with the finite thermal energy $\mathcal{U}%
_{T}$ in a finite volume and the finite spatial density $u_{T}$\ of thermal radiation.

\section{Contrasting Classical-Quantum Views for Radiation in a Minkowski
Frame}

\subsection{Normal Mode Expansions}

Both the classical and the quantum scalar field theories expand the fields in
terms of normal mode solutions of the scalar wave equation. \ For the
Minkowski vacuum situation in the classical case, we write the random
radiation field in a cubic box of side $L$ as a sum over all the linearly
independent normal mode solutions
\begin{equation}
\phi(ct,\mathbf{r)=}%
%TCIMACRO{\dsum \limits_{i}}%
%BeginExpansion
{\displaystyle\sum\limits_{i}}
%EndExpansion
\{a_{i}f_{i}(\mathbf{r)}\exp(-i\omega_{i}t)+a_{i}^{\ast}f_{i}^{\ast
}(\mathbf{r})\exp(i\omega_{i}t)\}
\end{equation}
where $f_{i}(\mathbf{r})\exp(-i\omega_{i}t)$ is a normalized solution of the
scalar wave equation, and $a_{i}=\exp(i\theta_{i})$ is a stochastic variable
associated with the random phase $\theta_{i}$ which is distributed randomly
over $[0,2\pi)$ and independently distributed for each wave solution $i$. The
complex conjugate $a_{i}^{\ast}=\exp(-i\theta_{i})$ involves the same random
phase $\theta_{i}$. \ For the case of vacuum (zero-point radiation), the
normalization for the solution $f_{i}(\mathbf{r})\exp(-i\omega_{i}t)$ is
chosen so that the radiation spectrum is Lorentz invariant and scale invariant
with an energy $(1/2)\hbar\omega$ per normal mode in the limit of unbounded
space. \ If the solutions $f_{i}(\mathbf{r})\exp(-i\omega_{i}t)$ are taken
over all Minkowski spacetime, then the field can be rewritten using
$a(\mathbf{k})=[L/(2\pi)]^{3/2}a_{i}$ and $(2\pi/L)^{3}\sum_{i}\rightarrow\int
d^{3}k$ as\cite{Coleak}
\begin{align}
\phi_{0}(ct,\mathbf{r)} &  \mathbf{=}%
%TCIMACRO{\dint }%
%BeginExpansion
{\displaystyle\int}
%EndExpansion
d^{3}k\frac{\mathfrak{h(}\omega)}{2}\{a(\mathbf{k)}\exp(i\mathbf{k\cdot
r}-i\omega t)+a(\mathbf{k)}^{\ast}\exp(-i\mathbf{k\cdot r}+i\omega
t)\}\nonumber\\
&  =%
%TCIMACRO{\dint }%
%BeginExpansion
{\displaystyle\int}
%EndExpansion
d^{3}k\mathfrak{h}(\omega)\cos[\mathbf{k\cdot r-\omega t+\theta(k})]
\end{align}
where%
\begin{equation}
\pi^{2}\mathfrak{h}^{2}(\omega)=\frac{1}{2}\frac{\hbar}{\omega}%
\end{equation}
When averaged over the random phases $\theta_{i}$, we find the average
values,
\begin{equation}
\left\langle a_{i}a_{j}\right\rangle =\left\langle \exp[i(\theta_{i}%
+\theta_{j})]\right\rangle =0=\left\langle a_{i}^{\ast}a_{j}^{\ast
}\right\rangle =\left\langle \exp[-i(\theta_{i}+\theta_{j})]\right\rangle
\end{equation}%
\begin{equation}
\left\langle a_{i}a_{j}^{\ast}\right\rangle =\left\langle a_{j}^{\ast}%
a_{i}\right\rangle =\left\langle \exp[i(\theta_{i}-\theta_{j})]\right\rangle
=\delta_{ij}%
\end{equation}

In the quantum case, we write the scalar field in a parallel fashion as
\begin{equation}
\overline{\phi}(ct,\mathbf{r)=}%
%TCIMACRO{\dsum \limits_{i}}%
%BeginExpansion
{\displaystyle\sum\limits_{i}}
%EndExpansion
\{\overline{a}_{i}f_{i}(\mathbf{r)}\exp(-i\omega_{i}t)+\overline{a}_{i}%
^{+}f_{i}^{\ast}(\mathbf{r})\exp(i\omega_{i}t)\}
\end{equation}
where here $\overline{a}_{i}$ is a quantum annihilation operator for the
quantum vacuum state $|0>$ while $\overline{a}_{i}^{+}$ is the associated
quantum creation operator. The quantum annihilation and creation operators
satisfy the commutation relations
\begin{equation}
\lbrack\overline{a}_{i},\overline{a}_{j}]=0=[\overline{a}_{i}^{+},\overline
{a}_{j}^{+}]
\end{equation}%
\begin{equation}
\lbrack\overline{a}_{i},\overline{a}_{j}^{+}]=\delta_{ij}%
\end{equation}
and have the vacuum expectation values%
\begin{equation}
\left\langle 0|\overline{a}_{i}\overline{a}_{j}|0\right\rangle =0=\left\langle
0|\overline{a}_{i}^{+}\overline{a}_{j}^{+}|0\right\rangle
\end{equation}%
\begin{equation}
\left\langle 0|\overline{a}_{i}\overline{a}_{j}^{+}|0\right\rangle
=\delta_{ij}%
\end{equation}
If the solutions $f_{i}(\mathbf{r})\exp(-i\omega_{i}t)$ are taken over all
Minkowski spacetime, then the quantum field $\overline{\phi}$ takes the form
\begin{equation}
\overline{\phi}(ct,\mathbf{r)=}%
%TCIMACRO{\dint }%
%BeginExpansion
{\displaystyle\int}
%EndExpansion
d^{3}k\frac{\mathfrak{h(}\omega)}{2}\{\overline{a}(\mathbf{k)}\exp
(i\mathbf{k\cdot r}-i\omega t)+\overline{a}^{+}(\mathbf{k)}\exp
(-i\mathbf{k\cdot r}+i\omega t)\}
\end{equation}
analogous to the first line of Eq. (6) for the classical case. \ However,
there is no quantum expression corresponding to the second line of Eq. (6)
because $\overline{a}(\mathbf{k})$ and $\overline{a}^{+}(\mathbf{k})$ in Eq.
(15)\ are operators rather than the complex numbers $a(\mathbf{k)}$ and
$a^{\ast}(\mathbf{k})$ appearing in Eq. (6).

\subsection{Vacuum Expectation Values of the Fields}

Despite the strikingly different points of view, the classical and quantum
scalar field theories give exact agreement between the (vacuum) two-point
field correlation function for classical fields and the symmetrized two-point
field vacuum expectation value for quantum fields. \ Thus simply using the
random-phase averages in Eqs. (8)-(9) for the classical fields and the vacuum
expectation values in Eqs. (13)-(14) for the quantum fields, we find%
\begin{align}
\left\langle \phi_{0}(ct,\mathbf{r)}\phi_{0}(ct^{\prime},\mathbf{r}^{\prime
})\right\rangle  &  =%
%TCIMACRO{\dsum \limits_{i}}%
%BeginExpansion
{\displaystyle\sum\limits_{i}}
%EndExpansion
\{f_{i}(\mathbf{r)}f_{i}^{\ast}(\mathbf{r}^{\prime})\exp[-i\omega
_{i}(t-t^{\prime})]+f_{i}^{\ast}(\mathbf{r)}f_{i}(\mathbf{r}^{\prime}%
)\exp[i\omega_{i}(t-t^{\prime})]\}\nonumber\\
&  =\left\langle 0|(1/2)\{\overline{\phi}(ct,\mathbf{r)}\overline{\phi
}(ct^{\prime},\mathbf{r}^{\prime})+\overline{\phi}(ct^{\prime},\mathbf{r}%
^{\prime})\overline{\phi}(ct,\mathbf{r)\}}|0\right\rangle
\end{align}
For the field expressions in all Minkowski spacetime given in Eqs. (6) and
(15), the correlation functions in Eq. (16) can be evaluated in closed form by
introducing a temporary cut-off at high frequency and then removing the
cut-off after the calculation. One finds\cite{conformal}%
\begin{align}
\left\langle \phi_{0}(ct,\mathbf{r)}\phi_{0}(ct^{\prime},\mathbf{r}^{\prime
})\right\rangle  &  =\left\langle 0|(1/2)\{\overline{\phi}(ct,\mathbf{r)}%
\overline{\phi}(ct^{\prime},\mathbf{r}^{\prime})+\overline{\phi}(ct^{\prime
},\mathbf{r}^{\prime})\overline{\phi}(ct,\mathbf{r)\}}|0\right\rangle
\nonumber\\
&  =\frac{\hbar c}{4\pi^{2}}%
%TCIMACRO{\dint }%
%BeginExpansion
{\displaystyle\int}
%EndExpansion
\frac{d^{3}k}{|k|}\cos[\mathbf{k}\cdot(\mathbf{r-r}^{\prime})-\omega
(t-t^{\prime})]\nonumber\\
&  =\frac{\hbar c}{2\pi|\mathbf{r}-\mathbf{r}^{\prime}|}%
%TCIMACRO{\dint \limits_{0}^{\infty}}%
%BeginExpansion
{\displaystyle\int\limits_{0}^{\infty}}
%EndExpansion
dk\nonumber\\
&  \times\{\sin[k(|\mathbf{r-r}^{\prime}|-c(t-t^{\prime})]+\sin
[k(|\mathbf{r-r}^{\prime}|+c(t-t^{\prime})]\}\nonumber\\
&  =\frac{-\hbar c}{\pi\lbrack c^{2}(t-t^{\prime})^{2}-(x-x^{\prime}%
)^{2}-(y-y^{\prime})^{2}-(z-z^{\prime})^{2}]}%
\end{align}
These correlation functions involve the inverse square of the
Lorentz-invariant proper time between the spacetime points $(ct,\mathbf{r)}$
and $(ct^{\prime},\mathbf{r}^{\prime})$ at which the fields are evaluated.
\ Since the metric of Minkowski spacetime is given by $ds^{2}=c^{2}%
dt^{2}-dx^{2}-dy^{2}-dz^{2},$ the coordinates $(ct,\mathbf{r)}$ are geodesic
coordinates, and the square of the distance along the geodesic between the two
spacetime point is exactly $[c^{2}(t-t^{\prime})^{2}-(x-x^{\prime}%
)^{2}-(y-y^{\prime})^{2}-(z-z^{\prime})^{2}].$ \ Thus equation (17)
corresponds exactly to the equation for zero-point radiation given in Eq. (4)
for a general spacetime. \ We notice that there is no distinguished
correlation length and no distinguished correlation time in this expression.
\ Higher-order correlation functions show a Gaussian behavior, and there is
complete agreement between the higher-order classical and quantum expressions
provided that the quantum operator order is completely
symmetrized.\cite{1975agree}

\subsection{Thermal Scalar Radiation}

Within classical theory with classical zero-point radiation, zero-point
radiation represents real radiation which is always present, and thermal
radiation is additional random radiation above the zero-point value. Thus if
$U(\omega,T)$ is the energy per normal mode at frequency $\omega$ and
temperature $T$, the thermal energy contribution is $U_{T}(\omega
,T)=U(\omega,T)-U(\omega,0).$ The thermal energy $\mathcal{U}_{T}(T)$ at
temperature $T$ in a box of finite size is finite and involves a finite
spatial density of thermal energy $u(T)=a_{Ss}T^{4}$ where $a_{Ss}$ is the
constant for scalar radiation corresponding to Stefan's constant for
electromagnetic radiation. The additional thermal energy is distributed across
the lower frequency modes of the radiation field, and it is the classical
zero-point radiation which prevents the thermal energy from leaking out to the
divergent spectrum of high frequency modes. Classical thermal radiation is
described in exactly the same random-phase fashion as the zero-point radiation
except that the spectrum $\mathfrak{h(\omega)}$ for scalar radiation takes the
form
\begin{equation}
\pi^{2}\mathfrak{h}^{2}(\omega)=U(\omega,T)c^{2}/\omega^{2}=[\hbar
c^{2}/(2\omega)]\coth[\hbar\omega/(2k_{B}T)]
\end{equation}
The calculation for the classical two-point field correlation function at
finite temperature accordingly takes exactly the same form as given above in
Eqs. (16)-(17), except that the spectrum is changed so that now
\begin{align}
\left\langle \phi_{T}(ct,\mathbf{r)}\phi_{T}(ct^{\prime},\mathbf{r}^{\prime
})\right\rangle  &  =\frac{\hbar c}{4\pi^{2}}%
%TCIMACRO{\dint }%
%BeginExpansion
{\displaystyle\int}
%EndExpansion
\frac{d^{3}k}{|k|}\coth\left[  \frac{\hbar ck}{2k_{B}T}\right]  \cos
[\mathbf{k}\cdot(\mathbf{r-r}^{\prime})-\omega(t-t^{\prime})]\nonumber\\
&  =\frac{\hbar c}{2\pi|\mathbf{r}-\mathbf{r}^{\prime}|}%
%TCIMACRO{\dint \limits_{0}^{\infty}}%
%BeginExpansion
{\displaystyle\int\limits_{0}^{\infty}}
%EndExpansion
dk\coth\left[  \frac{\hbar ck}{2k_{B}T}\right] \nonumber\\
&  \times\{\sin[k(|\mathbf{r-r}^{\prime}|-c(t-t^{\prime})]+\sin
[k(|\mathbf{r-r}^{\prime}|+c(t-t^{\prime})]\}
\end{align}

The quantum point of view regarding thermal radiation is strikingly different
from the classical viewpoint. The vacuum of the quantum scalar field is said
to involve fluctuations but no quanta, no elementary excitations, no scalar
photons, whereas the thermal radiation field involves a distinct pattern of
scalar photons. \ The quantum expectation values correspond to an incoherent
sum over the expectation values for the fields for all numbers $n_{\mathbf{k}%
}$ of photons of wave vector $\mathbf{k}$ with a weighting given by the
Boltzmann factor $\exp[-n_{\mathbf{k}}\hbar\omega_{\mathbf{k}}/(k_{B}T)].$
Thus the quantum two-point field correlation function is given
by\cite{1975agree}
\begin{align}
&  \left\langle |(1/2)\{\overline{\phi}(ct,\mathbf{r)}\overline{\phi
}(ct^{\prime},\mathbf{r}^{\prime})+\overline{\phi}(ct^{\prime},\mathbf{r}%
^{\prime})\overline{\phi}(ct,\mathbf{r)\}|}\right\rangle _{T}\nonumber\\
&  =%
%TCIMACRO{\dint }%
%BeginExpansion
{\displaystyle\int}
%EndExpansion
d^{3}k%
%TCIMACRO{\dsum \limits_{n=0}^{\infty}}%
%BeginExpansion
{\displaystyle\sum\limits_{n=0}^{\infty}}
%EndExpansion
\frac{1}{Z[\hbar ck/(k_{B}T)]}\exp\left[  \frac{-n_{\mathbf{k}}\hbar ck}%
{k_{B}T}\right] \nonumber\\
&  \times\left\langle n_{\mathbf{k}}|(1/2)\{\overline{\phi}(ct,\mathbf{r)}%
\overline{\phi}(ct^{\prime},\mathbf{r}^{\prime})+\overline{\phi}(ct^{\prime
},\mathbf{r}^{\prime})\overline{\phi}(ct,\mathbf{r)\}|}n_{\mathbf{k}%
}\right\rangle \nonumber\\
&  =\frac{\hbar c}{4\pi^{2}}%
%TCIMACRO{\dint }%
%BeginExpansion
{\displaystyle\int}
%EndExpansion
\frac{d^{3}k}{|k|}\coth\left[  \frac{\hbar ck}{2k_{B}T}\right]  \cos
[\mathbf{k}\cdot(\mathbf{r-r}^{\prime})-\omega(t-t^{\prime})]
\end{align}
where we have noted that
\begin{equation}
\frac{1}{2}\coth\frac{x}{2}=\frac{%
%TCIMACRO{\dsum \limits_{n=0}^{\infty}}%
%BeginExpansion
{\displaystyle\sum\limits_{n=0}^{\infty}}
%EndExpansion
(n+1/2)\exp[-nx]}{%
%TCIMACRO{\dsum \limits_{n=0}^{\infty}}%
%BeginExpansion
{\displaystyle\sum\limits_{n=0}^{\infty}}
%EndExpansion
\exp[-nx]}%
\end{equation}
and have defined%
\begin{equation}
Z(x)=%
%TCIMACRO{\dsum \limits_{n=0}^{\infty}}%
%BeginExpansion
{\displaystyle\sum\limits_{n=0}^{\infty}}
%EndExpansion
\exp[-nx]
\end{equation}
Thus for symmetrized products of quantum fields, the quantum expectation value
in Eq. (20) is in exact agreement with the corresponding classical average
value found in the first line of Eq. (19). \ Again the agreement holds for
higher order correlation functions provided the quantum operator order is
completely symmetrized.\cite{1975agree}

At a single spatial point $\mathbf{r\rightarrow r}^{\prime}$ but at two
different times $t$ and $t^{\prime}$, the classical and quantum correlation
functions (19) and (20) become\cite{thermBC}%
\begin{align}
\left\langle \phi_{T}(ct,\mathbf{r)}\phi_{T}(ct^{\prime},\mathbf{r}%
)\right\rangle  &  =\left\langle |(1/2)\{\overline{\phi}(ct,\mathbf{r)}%
\overline{\phi}(ct^{\prime},\mathbf{r})+\overline{\phi}(ct^{\prime}%
,\mathbf{r})\overline{\phi}(ct,\mathbf{r)\}|}\right\rangle _{T}\nonumber\\
&  =\frac{\hbar c}{2\pi}%
%TCIMACRO{\dint \limits_{0}^{\infty}}%
%BeginExpansion
{\displaystyle\int\limits_{0}^{\infty}}
%EndExpansion
dk\coth\left[  \frac{\hbar ck}{2k_{B}T}\right]  2k\cos[c(t-t^{\prime
})]\nonumber\\
&  =\frac{-\hbar}{\pi c}%
%TCIMACRO{\QOVERD{(}{)}{\pi k_{B}T}{\hbar}}%
%BeginExpansion
\genfrac{(}{)}{}{}{\pi k_{B}T}{\hbar}%
%EndExpansion
^{2}\frac{1}{\{\sinh[\pi k_{B}T(t-t^{\prime})/\hbar]\}^{2}}%
\end{align}

It should be emphasized again that although there is complete agreement
between the correlation functions arising in classical and quantum theories,
the interpretations in terms of fluctuations arising from classical wave
interference or in terms of fluctuations arising from the presence of photons
are completely different between the theories.\cite{1969} \ The contrast in
interpretations becomes even more striking when an accelerating coordinate
frame is involved.

\section{Contrasting Classical-Quantum Views for Radiation in a Rindler Frame}

\subsection{Rindler Coordinate Frame}

Although nonrelativistic physics allows a uniform gravitational field, this is
not possible in relativistic theory. The closest which one can come to the
nonrelativistic situation is that provided by the constant proper acceleration
of each point of a Rindler coordinate frame accelerating through Minkowski
spacetime. If the coordinates of a spacetime point in a Minkowski inertial
frame are given by $(ct,x,y,z)$, then the coordinates $(\eta,\xi,y,z)$ of the
Rindler frame which is at rest with respect to the rectangular coordinates at
time $t=0$ are given by%
\begin{equation}
ct=\xi\sinh\eta
\end{equation}%
\begin{equation}
x=\xi\cosh\eta
\end{equation}
with $y$ and $z$ retaining common values between the frames. A point with
constant spatial coordinates $(\xi,y,z)$ in the Rindler frame has coordinates
in the Minkowski frame given by $(x_{\xi},y,z)$ where $x_{\xi}$ changes with
time as
\begin{equation}
x_{\xi}=\xi\cosh\eta=(\xi^{2}+\xi^{2}\sinh\eta)^{1/2}=(\xi^{2}+c^{2}%
t^{2})^{1/2}%
\end{equation}
and so moves with acceleration $a_{\xi}=d^{2}x/dt^{2}=c^{2}/\xi$ at time
$t=0,$ and indeed in the Rindler frame has constant proper acceleration
\begin{equation}
a_{\xi}=c^{2}/\xi
\end{equation}
at all times. Thus for large coordinates $\xi$ the acceleration $a_{\xi}$
becomes small whereas for small $\xi$, the proper acceleration diverges. The
plane $\xi=0$ is termed the event horizon for the Rindler coordinate frame.

\subsection{Vacuum Correlation Functions in a Rindler Frame}

Scalar functions do not change their values under change of coordinates. Thus
we can write the scalar radiation fields $\phi_{R}(\eta,\xi,y,z)$ in the
Rindler coordinate frame as
\begin{equation}
\phi_{R}(\eta,\xi,y,z)=\phi(ct,x,y,z)=\phi(\xi\sinh\eta,\xi\cosh\eta,y,z)
\end{equation}
Since we have obtained the closed-form expression for the two-point field
correlation function for zero-point scalar fields in Eq. (17), it is easy to
rewrite the expression in terms of the Rindler coordinates to obtain
\begin{align}
&  \left\langle \phi_{R0}(\eta,\xi,y,z)\phi_{R0}(\eta^{\prime},\xi^{\prime
},y^{\prime},z^{\prime})\right\rangle \nonumber\\
&  =\left\langle \phi_{0}(\xi\sinh\eta,\xi\cosh\eta,y,z)\phi_{0}(\xi^{\prime
}\sinh\eta^{\prime},\xi^{\prime}\cosh\eta^{\prime},y^{\prime},z^{\prime
})\right\rangle \nonumber\\
&  =\frac{-\hbar c}{\pi}[(\xi\sinh\eta-\xi^{\prime}\sinh\eta^{\prime}%
)^{2}-(\xi\cosh\eta-\xi^{\prime}\cosh\eta^{\prime})^{2}\nonumber\\
&  -(y-y^{\prime})^{2}-(z-z^{\prime})^{2}]^{-1}\nonumber\\
&  =\frac{-\hbar c}{\pi\lbrack2\xi\xi^{\prime}\cosh(\eta-\eta^{\prime}%
)-\xi^{2}-\xi^{\prime2}-(y-y^{\prime})^{2}-(z-z^{\prime})^{2}]}%
\end{align}
We note that this correlation function depends upon only the time difference
$\eta-\eta^{\prime}$ and not upon the individual times $\eta$ and
$\eta^{\prime}.$ \ Thus zero-point radiation is a time-stationary distribution
of random radiation both in all inertial frames and in all Rindler frames. \ 

\subsection{"Thermal" Effects of Acceleration}

If one evaluates this expression (29) at a single spatial coordinate point
$(\xi,y,z)$ in the Rindler frame but at two different times $\eta$ and
$\eta^{\prime}$, then the two-time field correlation function becomes
\begin{align}
\left\langle \phi_{R0}(\eta,\xi,y,z)\phi_{R0}(\eta^{\prime},\xi
,y,z)\right\rangle  &  =\frac{-\hbar c}{\pi\lbrack2\xi^{2}\cosh(\eta
-\eta^{\prime})-2\xi^{2}]}\nonumber\\
&  =\frac{-\hbar c}{\pi\lbrack2\xi\sinh\{(\eta-\eta^{\prime})/2\}]^{2}%
}\nonumber\\
&  =\frac{-\hbar c}{\pi\lbrack2\xi\sinh\{[(\xi\eta-\xi\eta^{\prime
})/(2c)](c/\xi)\}]^{2}}\nonumber\\
&  =\frac{-\hbar(a_{\xi}/c)^{2}}{\pi c[2\sinh\{[(\tau_{\xi R}-\tau_{\xi
R}^{\prime})/2](a_{\xi}/c)\}]^{2}}%
\end{align}
where $\tau_{\xi R}=\xi\eta/c$ is the proper time recorded by a clock at rest
at horizontal coordinate $\xi$ in the Rindler frame. Written in this form, the
expression clearly involves a correlation time $\xi/c=c/a_{\xi}$ corresponding
to the time to travel the distance $\xi$ to the event horizon at speed $c$.
\ However, this correlation function also has exactly the same form as the
correlation function appearing in Eq. (23) corresponding to thermal radiation
with a Planck spectrum at a temperature

\begin{equation}
T_{\xi}=\frac{\hbar a_{\xi}}{2\pi ck_{B}}%
\end{equation}
where $a_{\xi}=c^{2}/\xi$ is the proper acceleration of a point at rest in the
Rindler frame at height $\xi.$ This temperature (31) is the
Unruh-Davies-Hawking temperature in quantum field theory.\ 

\subsection{The Classical Interpretation: Zero-Point Radiation}

Although both the \textit{classical} and \textit{quantum} correlation
functions take the same form (30) suggesting "thermal" behavior as seen in the
accelerating frame, the \textit{classical} interpretation still finds
zero-point radiation in the Rindler frame. \ Indeed if we consider the
two-point spatial correlation of the fields in a Rindler frame at a fixed time
$\eta=\eta^{\prime}$ but at two different spatial points, we find from Eq.
(29)%
\begin{align}
\left\langle \phi_{R0}(\eta,\xi,y,z)\phi_{R0}(\eta,\xi^{\prime},y^{\prime
},z^{\prime})\right\rangle  &  =\nonumber\\
&  =\frac{-\hbar c}{\pi\lbrack2\xi\xi^{\prime}\cosh(\eta-\eta)-\xi^{2}%
-\xi^{\prime2}-(y-y^{\prime})^{2}-(z-z^{\prime})^{2}]}\nonumber\\
&  =\frac{-\hbar c}{\pi\lbrack-(\xi-\xi^{\prime})^{2}-(y-y^{\prime}%
)^{2}-(z-z^{\prime})^{2}]}%
\end{align}
We see that the spatial correlation at a single time in the Rindler frame is
exactly that found in the Minkowski vacuum. There is no correlation length
whatsoever. Therefore no energy density above the zero-point radiation can be
defined and no classical thermal radiation can be present. \ 

\subsection{The Quantum Interpretation: Thermal Bath}

At the present time, \textit{quantum} theory has accepted statistical
mechanics as the foundation of thermodynamics, with the use of the classical
Boltzmann factor now being modified by the use of energy quanta. \ Thus for
quantum theory, "thermal" radiation involves a statistical sum such as appears
in equations (20) and (21). \ Indeed, in the review article by Crispino,
Higuchi, and Matsas, the authors check\cite{Crispino3} that a sum over quanta
found from a Bogolubov transformation from Minkowski over to Rindler space
indeed fits with the Planck spectrum found from the grand canonical ensemble
and from the KMS condition. \ Since the quantum ideas are found to fit with
the grand canonical ensemble, the authors conclude that the behavior is indeed
"thermal." \ On the other hand, Alsing and Milonni's derivation\cite{Milonni}
of the Planck factor involving the Fourier transform on acceleration through a
single plane wave apparently involves no randomness whatsoever.
\ Thermodynamic behavior without randomness seems surprising.

\subsection{Correlation Time Appearing from Acceleration Through Zero-Point
Radiation}

Although from Eq. (32) we found that there was no correlation length
associated with the zero-point correlation function seen in the Rindler frame,
from Eq. (30) we found there was indeed a correlation time $\xi/c$ which could
be associated with the acceleration $a_{\xi}=c^{2}/\xi.$ \ We wish to
emphasize that this correlation time is related to relativistic time behavior
in the Rindler frame and does not represent the label for a distinguished mode
which has energy above the zero-point energy. \ The correlation time $t_{a}$
associated with the Unruh-Davies-Hawking "temperature" corresponds to
$t_{a}=c/a=\xi/c$ which is the time for light to travel the distance to the
event horizon at speed $c.$ \ However, this correlation time represents merely
a relativistic time associated with a height $\xi$ in the Rindler frame and is
unrelated to thermodynamics. \ 

The correlation time $\xi/c$ found in Eq. (30) is imposed on the two-time
field correlation function for the vacuum situation by the trajectory through
spacetime of a point in the Rindler frame. \ Indeed the correlation function
involves exactly the geodesic length between spacetime points ($\eta,\xi,y,z)$
and $(\eta^{\prime},\xi,y,z).$ \ Since the corresponding geodesic coordinates
in the flat spacetime are ($ct=\xi\sinh\eta,x=\xi\cosh\eta,y,z)$ and
($ct^{\prime}=\xi\sinh\eta^{\prime},x=\xi\cosh\eta^{\prime},y,z),$ the
distance along the geodesic is given by $c^{2}(t-t^{\prime})-(x-x^{\prime
})=(\xi\sinh\eta-\xi\sinh\eta^{\prime})^{2}-(\xi\cosh\eta-\xi\cosh\eta
^{\prime})^{2}=-\xi^{2}[2-2\cosh(\eta-\eta^{\prime})]=[2\xi\sinh\{(\eta
-\eta^{\prime})/2\}]^{2}.$ \ This distance appears in the denominator of Eq.
(30) and agrees exactly with our definition of classical zero-point radiation
given in Eq. (4). \ Within classical physics, the preferred time has no
connection to any thermodynamic ensemble.

\subsection{Accelerating a Box of Classical Zero-Point Radiation}

A sense of the contrast between the classical and quantum points of view can
also be obtained by considering two sets of boxes of the same "large" size
with perfectly reflecting walls which keep all the radiation inside. \ The
boxes are chosen "large" in the sense that the surface effects are of
negligible importance compared to the intrinsic radiation correlation lengths
and times of interest. \ One set of boxes is always at rest in some inertial
frame and corresponds to the ensemble of classical zero-point radiation in a
Minkowski frame. \ The second set of boxes corresponds to a radiation ensemble
which is always at rest in the Rindler frame. At time $t=0=\eta,$ this second
set contains classical radiation identical to that in the first set of boxes
at rest in the inertial frame. \ 

In each set of boxes, the time evolution of the radiation must be obtained by
expanding the initial radiation pattern in terms of the normal modes for
radiation in the corresponding coordinate frame. \ The radiation normal modes
in the first set of boxes at rest in the Minkowski frame are different from
the radiation normal modes in the second set of boxes at rest in the Rindler
frame. \ However, as noted following Eq. (29), point, zero-point radiation is
time-stationary both in all inertial frames and in all Rindler frames. \ This
is the crucial point. The radiation modes may be different between the
inertial frame and the Rindler frame, but each frame contains a
time-stationary spectrum of random radiation which, at a single time in either
frame, agrees with the spatial distribution of radiation in the other frame.
\ Because the zero-point radiation is completely scale invariant and the
spectrum has no intrinsic correlation length whatsoever, the evolution of the
random radiation remains completely scale invariant. \ Thus if at some later
time $\eta$ in the Rindler frame, the time-evolved zero-point radiation in the
boxes at rest in the Rindler frame were compared with the zero-point radiation
in boxes of the same dimensions in the new inertial frame instantaneously at
rest with respect to the Rindler frame, there would be complete agreement
between the two ensembles. \ The Rindler frame perspective can introduce a
\textit{time} correlation associated with the acceleration as in Eq. (30), but
it can not introduce a \textit{spatial} correlation. \ The spatial radiation
pattern in the boxes accelerating with the Rindler frame remains zero-point
radiation. \ Only if there is some finite density of radiation above the
zero-point radiation is there the possibility of thermal equilibrium at
non-zero-temperature. \ Only in this case would the spatial correlations show
a variation in the energy density with the distance from the event horizon,
and only in this case would the radiation show the "sloshing" (change in the
relative position of the center of energy) of radiation if the box were
suddenly accelerated or the acceleration suddenly ceased. \ Indeed, "sloshing"
on acceleration seems a crucial sign of finite energy density within a box.
\ Zero-point radiation does not allow such "sloshing" because of its scale invariance.

\section{Closing Summary}

One speaks of the "thermal effects of acceleration" because of the appearance
of the correlation function associated with the Planck spectrum when a Rindler
coordinate system undergoes uniform acceleration through the zero-point
fluctuations in Minkowski spacetime. \ In this article we point out that there
is a disparity between the classical and quantum perspectives for this
phenomenon. \ Both the classical and the quantum fields $\phi$ and
$\overline{\phi}$ can be expanded in terms of the normal modes in the Rindler
coordinates. In the classical case, the random phases for the Rindler modes
can be reexpressed in terms of the random phases appearing in the Minkowski
modes. For the quantum fields, the Bogolubov transformation connects the
annihilation and creation operators of the Rindler modes to the annihilation
and creation operators of the Minkowski modes. In inertial frames, there is
agreement on the vacuum correlation functions between the classical and
quantum theories. \ Classical physics continues the special-relativistic
tensor behavior of the inertial frames into coordinate-change tensor behavior
for the Rindler frame, whereas quantum field theory introduces a new vacuum
state in the Rindler frame. \ The quantum analysis looks at the two-time
correlation function and notes the appearance of the Planck spectrum without
considering the associated spatial correlations of the fields. Because the
correlation function in time can be associated with the canonical ensemble,
the quantum literature refers to "thermal" behavior. \ On the other hand,
because the relativistic classical point of view does not define thermal
behavior in terms of a canonical ensemble, there is much less willingness to
identify the relativistic radiation in the Rindler frame as "thermal"
radiation. \ Indeed the relativistic classical point of view insists that the
scale-invariant vacuum state is unique, involves tensor behavior between
coordinate frames, and can not be redefined for different coordinate frames.
\ The distance along a geodesic between two spacetime points is an invariant,
despite the varying appearance in different coordinate frames. \ In classical
field theory, there is nothing comparable to the quantum distinction between
the Minkowski vacuum state and the Rindler vacuum state. \ The classical view
suggests that the effects of acceleration through zero-point radiation are not
thermal but rather are associated with time correlations imposed on the
scale-invariant zero-point radiation due to the parameters of the trajectory
through spacetime. \ The classical viewpoint suggests that an accelerating
thermometer will not record an elevated temperature.

Within a relativistic classical radiation theory, we expect thermal radiation
to be strongly associated with ideas of relativity. \ Indeed, the zero-point
correlations can be linked to thermal correlations when finite amounts of
additional radiation are introduced. By insisting that there is but one
correlation time in a Rindler frame involving classical thermal radiation, one
can obtain a derivation of the Planck spectrum for relativistic classical
thermal radiation in a Minkowski frame.\cite{ScalarRad}

\section{Note Added in Manuscript}

This article has received sharp criticism from referees who are strongly
antagonistic to its point of view. \ It has been suggested that the article
fails to recognize the "fact" that accelerating objects indeed experience
elevated temperatures, "Steaks will cook, eggs will fry." \ Now this is a
"fact" for which there is no experimental evidence. \ The present analysis
indeed suggests that this idea may be an error. \ Criticism has also been
directed to the article's failure to discuss "detectors" and the focus upon
merely the radiation present in the Rindler frame. \ However, it is one of the
most fundamental ideas of thermodynamics that, in equilibrium, a dectector and
the radiation at the same spatial point will be at the same temperature.
\ Thus we should be able to determine the temperature of any "detector" by
investigating the temperature of the radiation with which it is in
equilibrium. \ Criticism has also been directed to the fact that the classical
radiation discussed is not retained within reflecting boundaries. \ However,
this criticism also seems without merit. \ As seen in the Rindler frame, the
zero-point radiation correlation function of Eq. (29) is stationary in time
(involving only time differences $\eta-\eta^{\prime})$, and hence the random
radiation in the Rindler frame can be expressed in terms of the radiation
normal modes of the Rindler frame with random phases between the normal modes.
\ If conductors are introduced to provide a finite-length box for the
radiation modes, then the correlation function will be altered only at the
low-frequency modes near the fundamental associated with the finite length of
the box. \ As the box becomes increasingly large, the correlation function
will go over to the free-space expression given in Eq. (29) for which the
analysis was given. \ Thus the finite length of the box and the presence of
accelerating mirrors should not change the arguments of the present article.
\ It is noteworthy that the major quantum field theory literature, including
the original work by Davies and the review article by Crispino et al., makes
no use of finite-sized boxes in the analysis of the thermal effects of
acceleration. \

\end{document}